# Theoretical prediction of the half-metallicity in one-dimensional $Cr_2NO_2$ nanoribbons

Guo Wang*, Yi Liao

Department of Chemistry, Capital Normal University, Beijing 100048, China

*Corresponding author. Email: wangguo@mail.cnu.edu.cn

ABSTRACT

One-dimensional $Cr_2NO_2$ nanoribbons cutting from the oxygen-passivated $Cr_2NO_2$ MXene are investigated by using density functional theory. The wide nanoribbons have ferromagnetic ground states and are half-metals, independent of their chirality. The half-metallic band gaps of the wide nanoribbons are larger than 1 eV, which are large enough for avoiding thermally activated spin flip. The magnetism does not rely on the edge states but originates from all the Cr atoms. Furthermore, the half-metallicity is still robust in an electronic device even if the bias is up to 1 V. Therefore, one-dimensional $Cr_2NO_2$ nanoribbons are good candidates for spintronics.

## 1. Introduction

A half-metal [1], in which a spin channel is metallic while the other is insulating, is a novel material that has promising applications in spintronics. By utilizing this kind of materials, spins instead of charges in electronic devices are manipulated so that the power consumption can be significantly decreased. With the realization of single-layer two-dimensional graphene [2], the size of an electronic device can be greatly reduced. A low-dimensional half-metal is extremely attractive for realizing electronic devices. Zigzag graphene nanoribbons belong to such a kind of materials. They were predicted to be half-metallic [3]. However, the magnetism and half-metallicity come from the edge states, which are sensitive to the edge atomic structures. Furthermore, the half-metallicity requires strong external electric field that may hinder its application. Intrinsic half-metallicity is highly desirable, which does not require any external condition. Recently, intrinsic two-dimensional half-metals with atomic thickness, such as g-$C_4N_3$ [4], $Co_9Se_8$ [5], $FeCl_2$ [6], $Cr_2C$ [7], $YN_2$ [8], $TiCl_3$ [9], $VCl_3$ [9, 10], $VI_3$ [10], $Cr_2NO_2$ [11] and $FeC_2$ [12], were predicted. The intrinsic half-metallicity prevents defect or vacancy controlling, and does not require doping with foreign transition metal atoms that may form clusters [9]. Thus, these two-dimensional structures are promising candidates for spintronics. In electronic devices, the current from source to drain is in one-dimension, so one-dimensional materials are enough for building the device and should further reduce its size.

One-dimensional structures, such as BN [13] and GaS nanoribbons [14], were predicted to be half-metallic. However, similar to the zigzag graphene nanoribbons [3], the half-metallicity also originates from the edge states, which may be affected by foreign atoms [9]. It is desirable to design a one-dimensional half-metal that does not only rely on the edge states. It is noteworthy that the two-dimensional $Cr_2NO_2$ MXene is an oxygen-passivated half-metal [11] that should be stable against oxidization under atmosphere condition. In the present work, one-dimensional $Cr_2NO_2$ nanoribbons cutting from the two-dimensional $Cr_2NO_2$ MXene were investigated by using density functional theory. Unlike the zigzag graphene nanoribbons, the half-metallicity of the one-dimensional $Cr_2NO_2$ nanoribbons originates from all the Cr atoms. Furthermore, the half-metallicity is still robust in an electronic device even if the bias is up to 1 V.

## 2. Models and computational details

Both armchair and zigzag $Cr_2NO_2$ nanoribbons cutting from the two-dimensional $Cr_2NO_2$ MXene were investigated. The top and side views of an armchair $Cr_2NO_2$ nanoribbon with width of eleven atoms (abbreviated as 11-ACrNR) are shown in Figure 1(a) and 1(c). For zigzag type, the situation is more complex [15]. A zigzag $Cr_2NO_2$ nanoribbon with width of twenty one atoms (21-ZCrNR) is shown in Figure 1(b) and 1(d).

However, there could be another structure with different edges. Since the O atoms are always directed to the Cr atoms in the opposite layer [16], only the arrangement of the Cr and N atoms is needed to distinguish them. From the bottom to the top, the orders of the atoms for 21-ZCrNR in Figure 1(b) are both "CrCrN" at the two edges. If the orders are both "CrNCr", the nanoribbon is called 21-Z'CrNR. When the width $N$ equals to $3p$ where $p$ is a positive integer, the situation is the same and the nanoribbons are defined as ZCrNRs and Z'CrNRs accordingly. When $N$ equals to $3p+1$, the nanoribbons with "CrCrN" arrangement at an edge and "CrNCr" at the other edge are defined as ZCrNRs, while those with "NCrCr" and "CrCrN" are defined as Z'CrNRs. When $N$ equals to $3p+2$, the nanoribbons with "CrCrN" and "NCrCr" are defined as ZCrNRs, while those with "CrNCr" and "CrCrN" are defined as Z'CrNRs.

The primitive cells of the ACrNRs with width from 5 to 30 atoms as well as those of the ZCrNRs and Z'CrNRs with width from 5 to 37 were optimized by using the VASP program [17]. The PBE density functional [18] and projector-augmented wave with cutoff energy 400 eV were adopted. Spin polarization was considered in the calculations. For Cr atoms with strongly correlated 3d electrons, the GGA+U method was used. The difference between the onsite coulomb and exchange parameters was set to 3 eV, which was successfully applied to $CrO_2$ [19], $Cr_2C$ MXene [7] and $Cr_2N$ MXene [11]. A Monkhorst-Pack sampling with 41 k-points along the periodic direction in the first Brillouin zone and vacuum with 15 Å spacing along the non-periodic directions were used.

## 3. Results and discussions

In the ferromagnetic (FM) solutions, all the nanoribbons are half-metals. The band structures of 21-ZCrNR are shown in Figure 2. Like the band structures of the other nanoribbons, there are many bands going across the Fermi level for majority spin as shown in Figure 2(a) while there is a band gap for minority spin in Figure 2(b). The nanoribbons are metallic for majority spin while insulating for minority spin. The band gaps for minority spin are shown in Figure 3(a). In the figure, the band gaps of the ACrNRs decrease generally with the width. Small oscillation occurs especially when $N$ is less than 21. The band gap of 21-ACrNR is 2.39 eV and this value changes little when $N$ is larger than 21. For the ZCrNRs, the situation is similar but the oscillation is more obvious when $N$ is less than 21. The band gap of 21-ZCrNR is 1.65 eV. The band gaps of all the ZCrNRs are smaller than those of the ACrNRs except for $N=7$. Unlike the other two types of the nanoribbons, the band gaps of the Z'CrNRs do not generally decrease with the width but strongly oscillate. The band gaps for the Z'CrNRs with $N=3p+1$ are much larger than those of the others. For example, the band gap of 22-Z'CrNR is 2.70 eV, which is larger than the values 1.93 and 1.99 eV for 21-Z'CrNR and 23-Z'CrNR, respectively. The larger band gaps of Z'CrNRs with $N=3p+1$ may come from their special atomic arrangement with N atoms at the both edges.

The half-metallic band gap, defined as the difference between the Fermi level and the valence band maximum [13], is also calculated for the nanoribbons. As shown in Figure 3(b), the half-metallic band gaps are all smaller than the corresponding band gaps for minority spin. This is reasonable according to the definition. The change of the band gaps is quite small for the ACrNRs with $N>6$ and ZCrNRs with $N>10$. The half-metallic band gaps are 1.87 and 1.01 eV for 7-ACrNR and 11-ZCrNR, respectively. For Z'CrNRs, the oscillation also occurs for the half-metallic band gaps and the values for structures with $N=3p+1$ are also the largest. The half-metallic band gaps are in the range of 1.17-2.31 eV when $N>6$. The half-metallic band gaps of all the nanoribbons are larger than 1 eV when $N>8$. These values are large enough for avoiding thermally induced spin flip and should benefit their applications in spintronics.

The spin density of 21-ZCrNR is shown in Figure 4. From the figure, it can be seen that the magnetic moments of 21-ZCrNR come from all the Cr atoms, not only from the edge atoms [13, 14]. For the other nanoribbons, the situation is similar. The magnetic moments per Cr atom are 2.5 $\mu_B$ for all the ACrNRs. For ZCrNRs and Z'CrNRs with different edges, the magnetic moments are in the range of 2.5-3.25 and 1.75-2.5 $\mu_B$, respectively. The magnetism does not only originate from the edge states, and should be more realizable in experiments.

Besides the FM configuration, anti-ferromagnetic (AFM) configurations were also calculated to verify the ground states. Since the spins in the zigzag graphene nanoribbons couple anti-ferromagnetically between the two edges [3], the configurations with anti-ferromagnetic coupling between the top half and the bottom half in Figure 1(a) and 1(b) were also calculated (denoted as AFM1-1). Because the nanoribbons are structures with N atoms sandwiched between two Cr layers, configurations with anti-ferromagnetic coupling between the two layers are possible (AFM1-2). Moreover, configurations with anti-ferromagnetic coupling between the left half and the right half (AFM1-3), as well as those with the couplings both occurred in AFM1-2 and AFM1-3 (denoted as AFM1-4) were calculated. Besides the above four AFM configurations that are based on the primitive cells, configurations with anti-ferromagnetic coupling between the two adjacent cells were also calculated (AFM2). Because the band gap changes little when $N>21$, the AFM configurations of the nanoribbons with $N$ no more than 21 are calculated. The relative energies per Cr atom of the AFM configurations with respect to the corresponding FM configurations are shown in Figure 5. In Figure 5(a), all the energies of the AFM configurations of the ACrNRs are positive, indicating that they are less stable than the FM configurations and the FM configurations are the ground states. In Figure 5(b), some AFM configurations of the ZCrNRs are more stable than the FM configurations. For 8-ZCrNRs, all the five AFM configurations are more stable than the FM configurations. The AFM2 configuration is the ground state. The AFM2 configurations of 5-ZCrNR, 6-ZCrNR, 10-ZCrNR, 11-ZCrNR, 12-ZCrNR and 17-ZCrNR are also the ground states. For 10-ZCrNR, 12-ZCrNR and 17-ZCrNR, the energy differences between the FM and AFM2 configurations are only 4, 1 and 4 meV. For the wider ZCrNRs, the FM configurations are the ground states. As for the Z'CrNRs, the FM configurations become the ground states when $N>6$. The results for the AFM configurations ensure that the ground states of the wide nanoribbons are the FM configurations. This guarantees their half-metallicity.

The relative stabilities between different types of all the nanoribbons are difficult to determine, due to their different stoichiometries. All the ACrNRs have a chemical formula $Cr_2NO_2$, while this only occurs for ZCrNRs and Z'CrNRs when $N=3p$. The other nanoribbons all have slightly different chemical formulas because of their different edges. The 5-ACrNR, 15-ZCrNR and 15-Z'CrNR all have ten Cr atoms. The 6-ACrNR, 18-ZCrNR and 18-Z'CrNR, as well as 7-ACrNR, 21-ZCrNR and 21-Z'CrNR, share the same numbers of Cr atoms (twelve and fourteen). Furthermore, they all have the same magnetic moment per Cr atom (2.5 $\mu_B$). Compared with the corresponding ACrNRs, the relative energies per Cr atom of 15-ZCrNR, 18-ZCrNR and 21-ZCrNR are -184, -131 and -119 meV, respectively, while the relative energies of the Z'CrNRs are 37, 41 and 26 meV, respectively. For the above nine nanoribbons, the ZCrNRs are the most stable, while the Z'CrNRs are the least stable.

The 21-ZCrNR is more stable than its counterparts (7-ACrNR and 21-Z'CrNR) and it shares common band gap of the wide ZCrNRs. These are reasons why it was taken as an example to present the band structures in Figure 2. The half-metallic band gap (1.07 eV) is much larger than the thermal energy, and therefore possible thermally activated spin flip is prevented. However, the models of the above nanoribbon are infinitely long structures. It is important to make clear whether the half-metallicity is sustainable in very short structures. For this reason, a nanometer-sized device was also investigated based on the non-equilibrium Green's function method. The current-bias ($I$-$V$) characteristics were obtained by using the OPENMX program [20] under the Landauer-Büttiker formula [21]. The stable 21-ZCrNR was also used as an example. The left and right semi-infinite electrodes are composed of two primitive cells, while the central scattering region was built with three primitive cells. The GGA+U method similar to that used for the infinitely long structures was adopted. Norm-conserving pseudopotentials [22] and variationally optimized pseudo-atomic localized basis functions [23, 24] were used. Each s, p, or d (for Cr atoms only) orbital was described by a primitive orbital. A sampling with 101 k-points along the one-dimensional transport direction was used. In Figure 6, the total current or the current for majority spin becomes non-zero at a very small bias 0.1 V and the current increases with the bias. The current is tens of μA and the current for minority spin is always zero even if the bias is up to 1 V. The spin polarization is 100% in this bias region. These indicate that the nanometer-sized nanoribbon is also half-metallic. Therefore, the one-dimensional nanoribbon should be a good candidate for spintronics.

## 4. Conclusions

Several armchair and zigzag one-dimensional $Cr_2NO_2$ nanoribbons cutting from the oxygen-passivated stable $Cr_2NO_2$ MXene are investigated by using density functional theory. It is indicated that the FM configurations are the ground states except for a few narrow nanoribbons, which have AFM2 ground states. All the nanoribbons with FM ground states are half-metals, independent of their chirality. In their band structures, there are many bands going across the Fermi level for majority spin while there is a band gap for minority spin. The band gaps for minority spin and the half-metallic band gaps oscillate with the width for the Z'CrNRs. However, the band gaps generally decrease with the width for the other nanoribbons and then change little when the width is above a certain threshold. The half-metallic band gaps of all the nanoribbons are larger than 1 eV when $N>8$. These values are large enough for avoiding thermally activated spin flip. Spin density analysis indicates that the magnetism comes from all the Cr atoms. This is different from BN and GaS nanoribbons, whose magnetism relies on the edge states. The magnetism and the half-metallicity should be more applicable in real application. For some nanoribbons with the same chemical formula $Cr_2NO_2$, the ZCrNRs are the most stable while the Z'CrNRs are the least. The half-metallicity is still robust in a device composed of 21-ZCrNR even if the bias is up to 1 V. Therefore, the one-dimensional $Cr_2NO_2$ nanoribbons should be good candidates for spintronics.

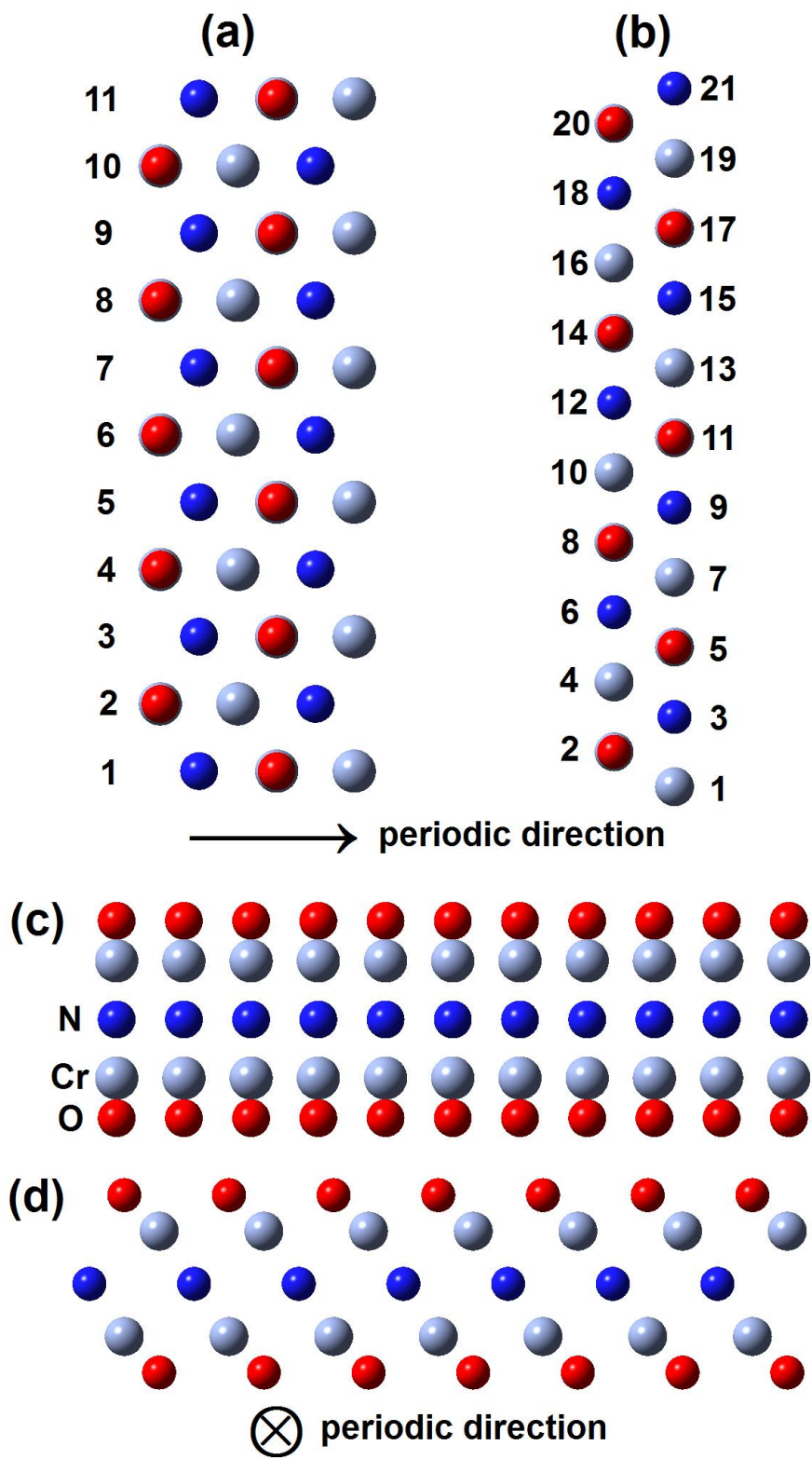

**Figure 1.** Top view of the primitive cells of (a) 11-ACrNR, (b) 21-ZCrNR and side view of the primitive cells of (c) 11-ACrNR, (d) 21-ZCrNR.

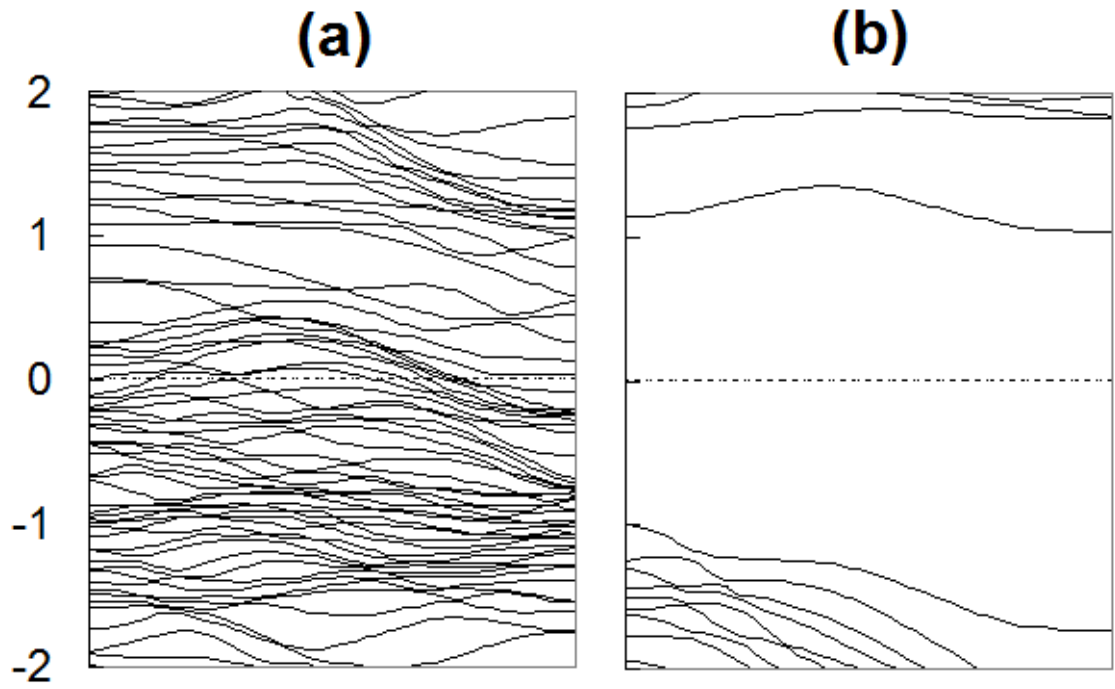

**Figure 2.** Band structures of 21-ZCrNR for (a) majority and (b) minority spins. Horizontal axis, reciprocal lattice vector from Γ to X; vertical axis, energy in eV. Fermi level is set to 0.

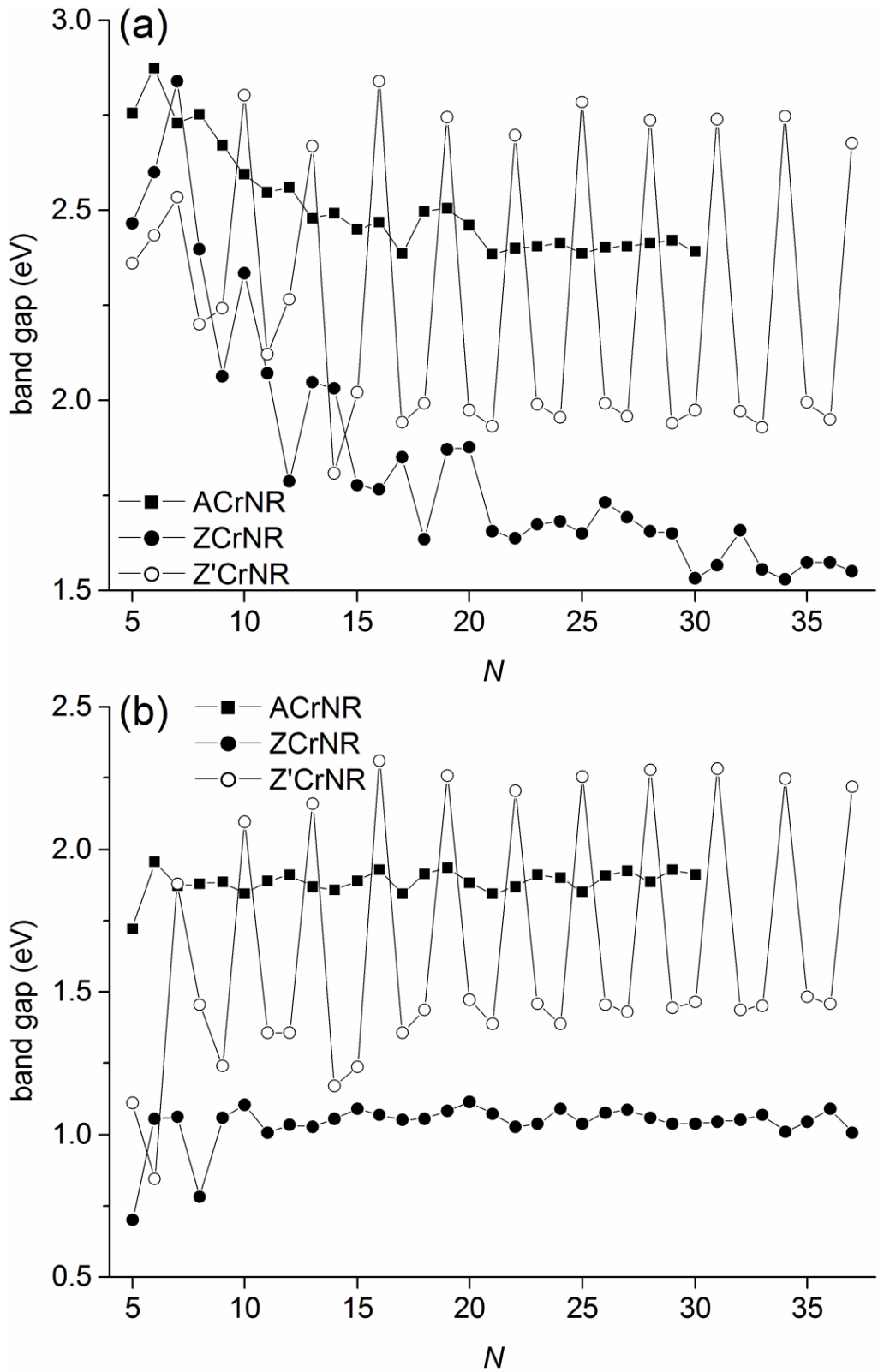

**Figure 3.** (a) Band gaps for minority spin and (b) half-metallic band gaps.

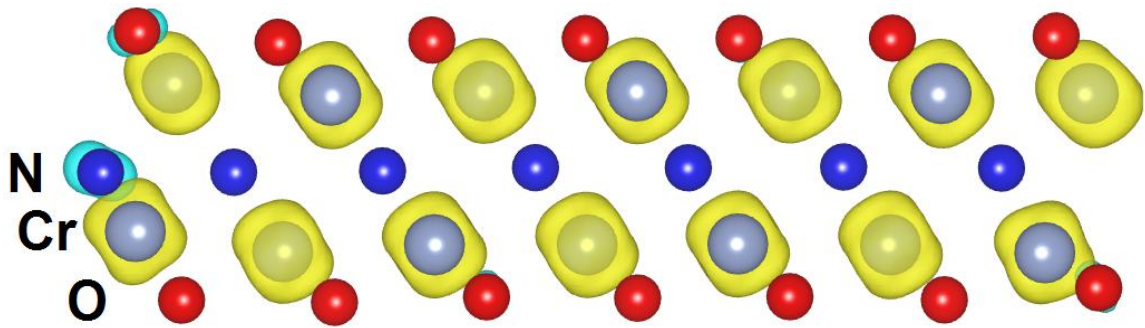

**Figure 4.** Spin density of 21-ZCrNR.

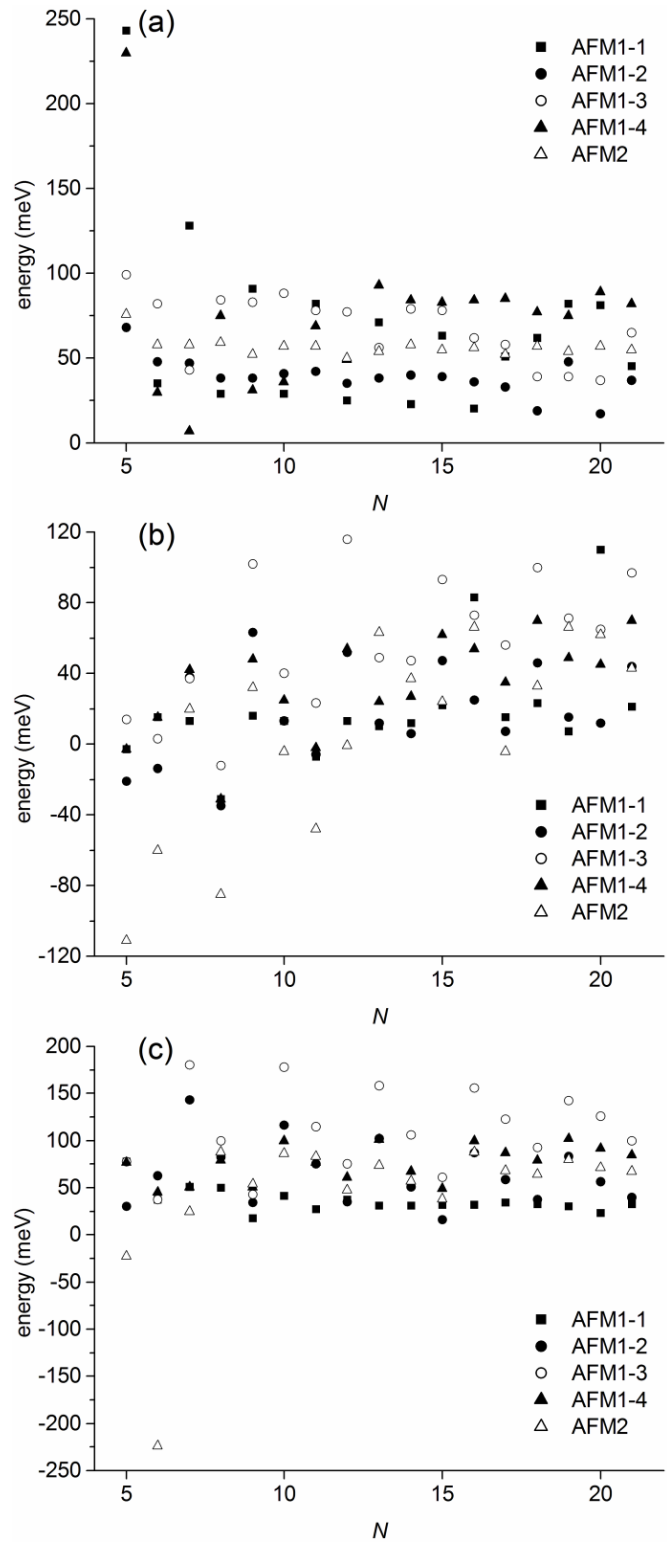

**Figure 5.** Relative energies per Cr atom of AFM configurations with respective to FM configurations for (a) ACrNRs, (b) ZCrNRs and (c) Z'CrNRs.

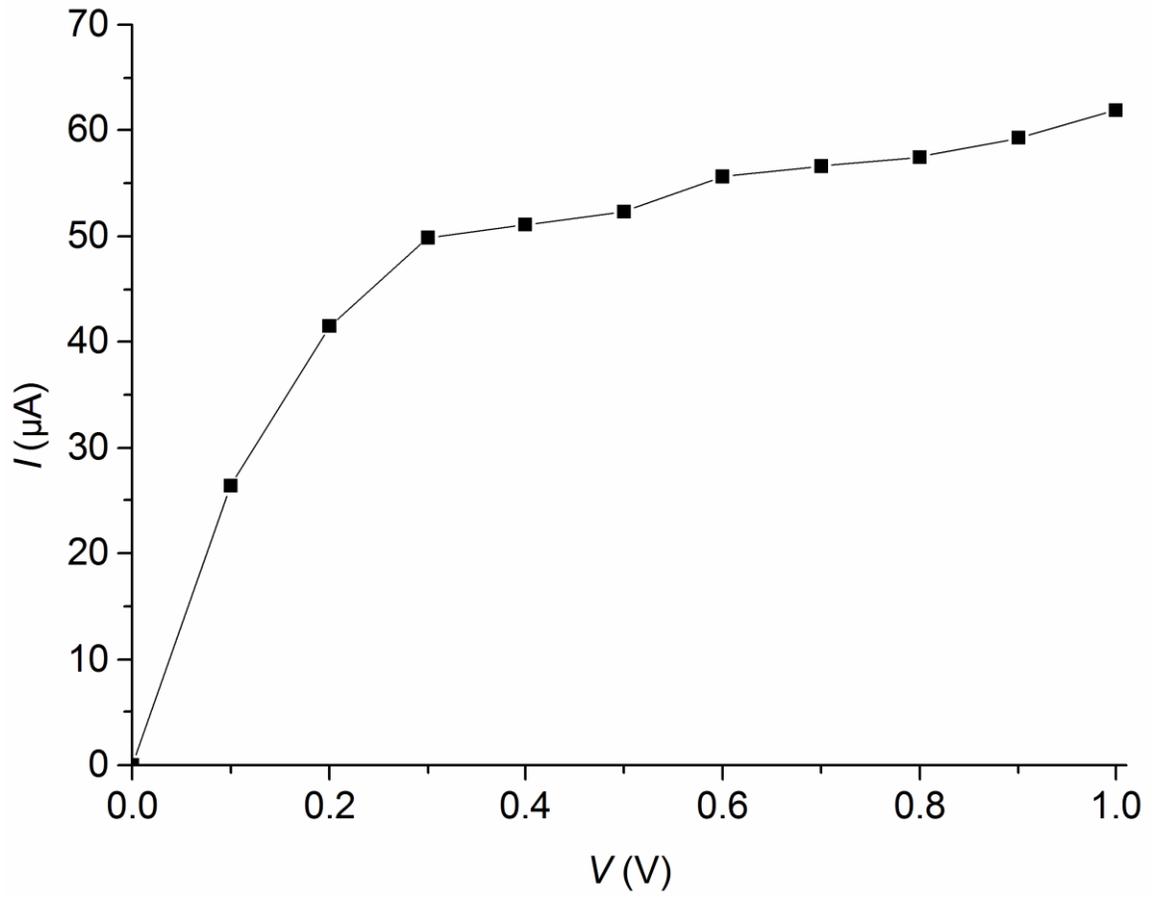

**Figure 6.** *I-V* characteristics of electronic device composed of 21-ZCrNR.